\documentclass{DISproc}

\usepackage{graphicx}
\usepackage{wrapfig}
\usepackage{rotating}
\usepackage{amssymb, amsmath, array}
\usepackage{hepunits}
\usepackage{csquotes}
\usepackage{booktabs}
\usepackage{hyperref}

\newcommand{\pom}{{I\!\!P}}
\newcommand{\reg}{{I\!\!R}}

\newcommand{\mx}{M_{_{\rm X}}}

\begin{document}
\title{Revisiting QCD Fits in Diffractive DIS}

\author{{\slshape Federico A.\ Ceccopieri$^{1,2}$, Laurent Favart$^2$}\\[1ex]
$^1$IFPA, Universit\'e de Li\`ege,  All\'ee du 6 ao\^ut, B\^at B5a, 4000 Li\`ege, Belgium.\\
$^2$Universit\'e Libre de Bruxelles, Boulevard du Triomphe, 1050 Bruxelles, Belgium.}

\contribID{282}
\doi  
\maketitle

\begin{abstract}
A new method of extracting diffractive parton distributions is presented which avoids 
the use of Regge theory ansatz and is in much closer relation with the factorisation theorem 
for diffractive hard processes.
\end{abstract}

\section{Introduction}
\noindent\\
Diffractive parton distributions functions (DPDF's)~\cite{FF} are essential ingredients in the understanding and 
description of hard diffractive processes. 
The factorisation theorem for diffractive Deep Inelastic Scattering (DIS)~\cite{DDISfactorisation} enables one to factorise the diffractive DIS cross-section into a 
long-distance contribution, parametrised by DPDF's, from a short-distance one, pertubatively calculable. 
Although DPDF's encode non-perturbative effects of QCD dynamics and therefore must be extracted from data,  
their dependence on the factorisation scale is predicted by pQCD~\cite{FF}.  
Moreover the short distance cross-section is the same as inclusive DIS~\cite{DDISfactorisation} 
so that higher order corrections can be systematically accounted for.
Due to the factorisation theorem, DPDF's are universal distributions
in the context of diffractive DIS and diffractive dijet cross-sections
are well described by next-to-leading order predictions based on DPDF's~\cite{diffdijet}.
The commonly used approach~\cite{diffdijet,H1LRG06,Zeus_combo,recent_global_fit} 
to extract DPDF's is to assume proton vertex factorisation, \textsl{i.e.} that DPDF's can be factorised into a flux factor depending only on $x_\pom$ and $t$ and a term depending only 
on the lepton variables $\beta$ and $Q^2$: 
\begin{equation}
f_i^D (\beta, Q^2 , x_\pom , t) = f_{\pom/P} (x_\pom , t) \; f_i^{\pom} (\beta, Q^2 ) 
+ f_{\reg/P} (x_\pom , t) \; f_i^{\reg} (\beta, Q^2 ) + ...\nonumber
\end{equation}
Each term in the expansion, according to Regge theory, is supposed to give a dominant contribution in a given range of $x_\pom$, the pomeron ($\pom$) at low $x_\pom$, the reggeon ($\reg$) at higher value of $x_\pom$ and so on. The flux factor  $f_{\pom/P}$ ($f_{\reg/P}$) can be interpreted as the probability that a pomeron (reggeon) with a given value of $x_\pom$ and $t$ couples to the proton. This approach assumes an arbitrary truncation of the trajectory expansion 
and requires that parton distributions of each trajectory ($f_i^{\pom}$, $f_i^{\reg}$, ...)
should be simultaneously extracted from data. It therefore introduces a large number of parameters in the fit 
and it is potentially biased by the choices of the flux factors. Although it has been proven 
to be supported by phenomenological analyses within HERA-I data precision, it is not 
routed in perturbative QCD and might be not entirely satisfactory with the expected precision
increase of HERA-II data.

\section{The new method}
\begin{figure}[t]
\begin{center}
\begin{minipage}{16pc}
\includegraphics[width=15pc]{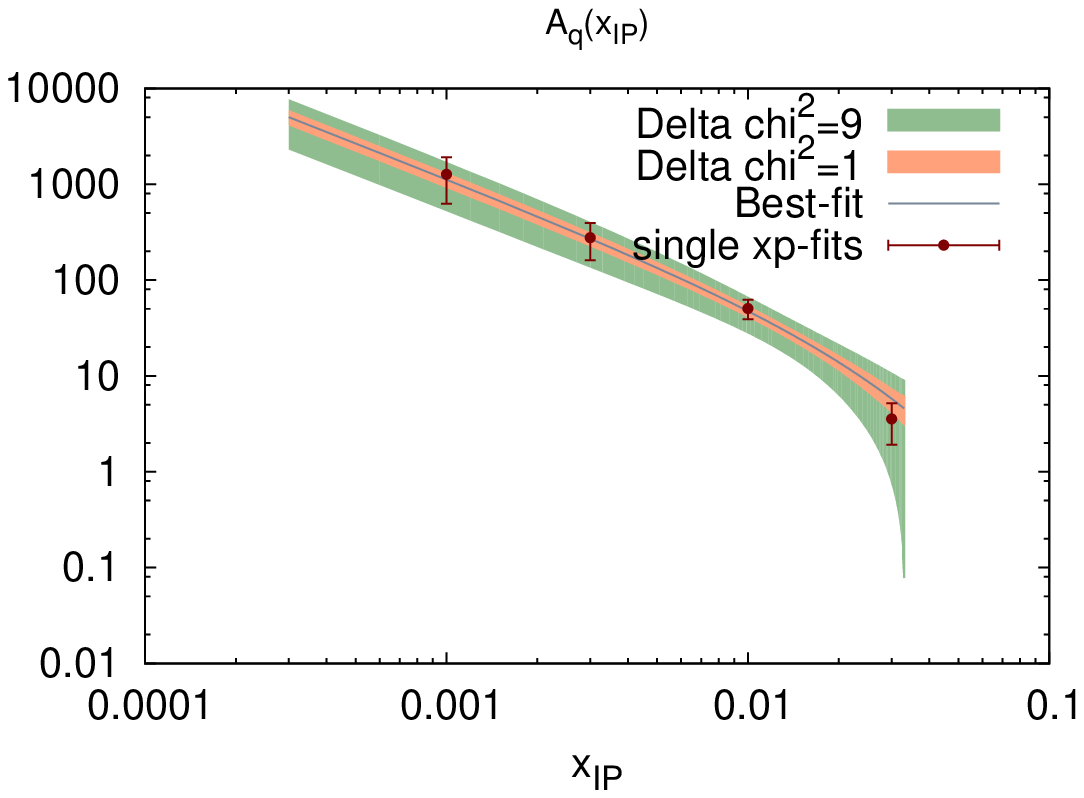}
\end{minipage}
\begin{minipage}{16pc}
\includegraphics[width=15pc]{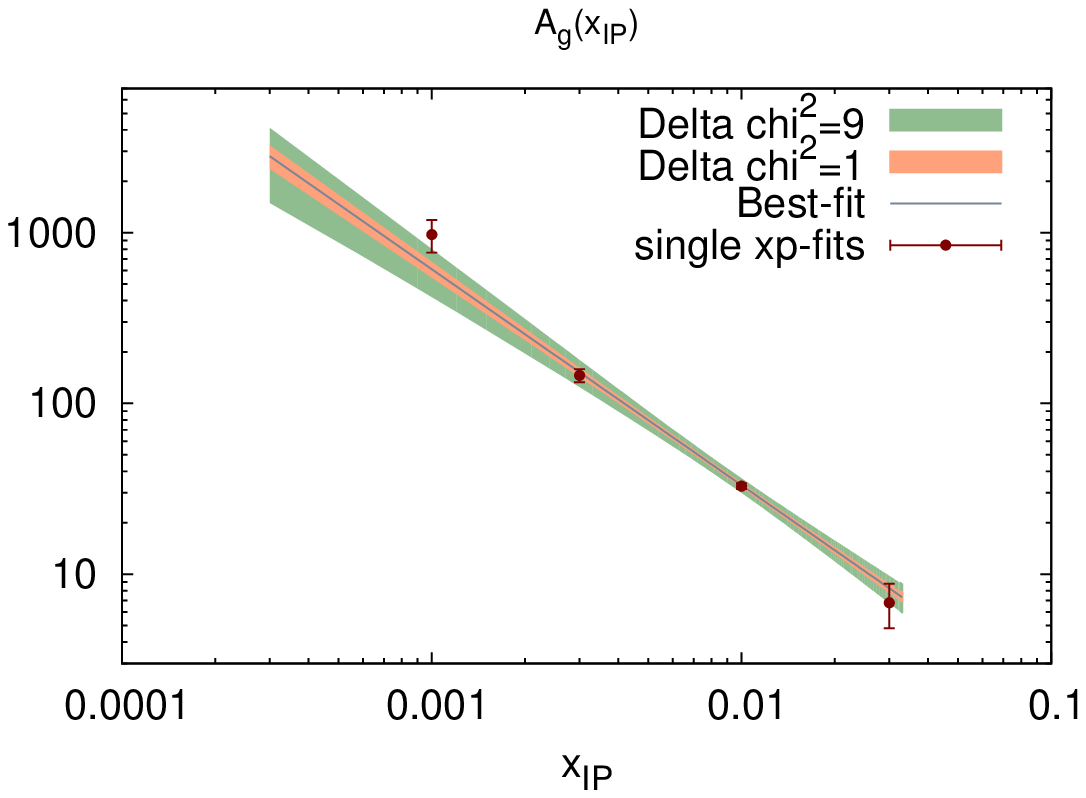}\\
\vspace*{-0.5cm}
\end{minipage} 
\begin{minipage}{16pc}
\hspace{0.5pc}
\includegraphics[width=14.5pc]{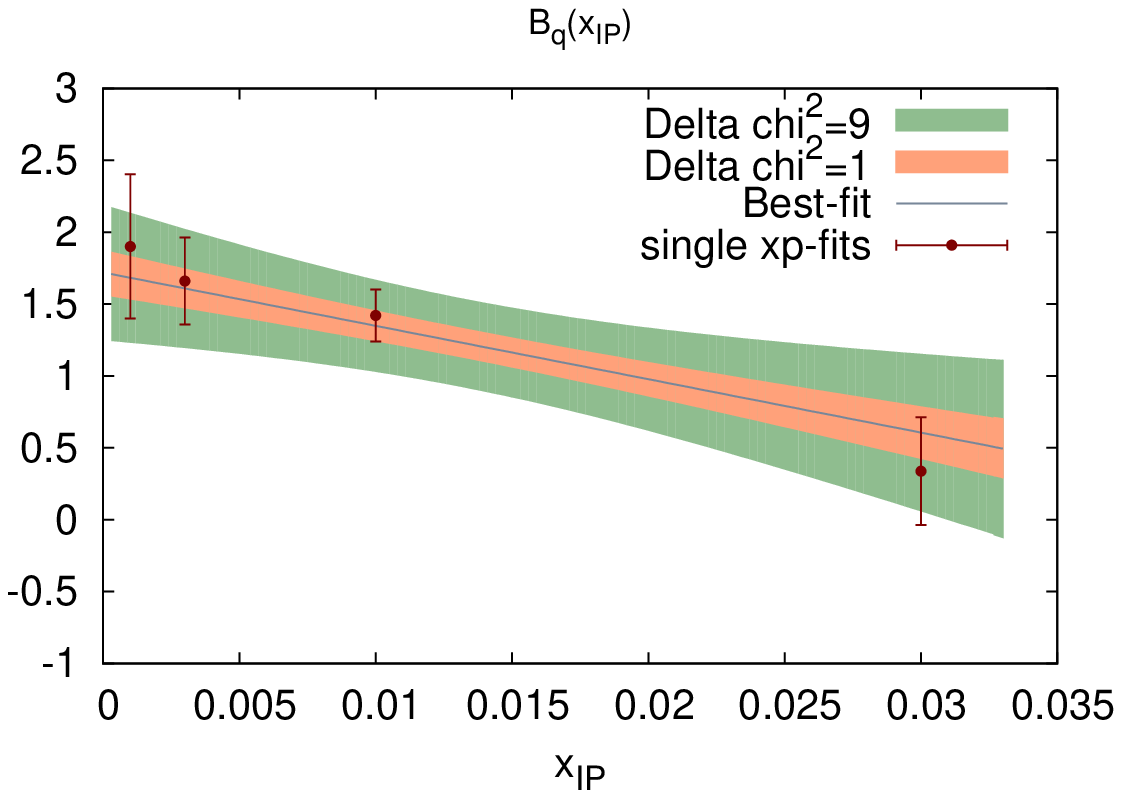}
\end{minipage}
\begin{minipage}{16pc}
\hspace{0.5pc}
\includegraphics[width=14.5pc]{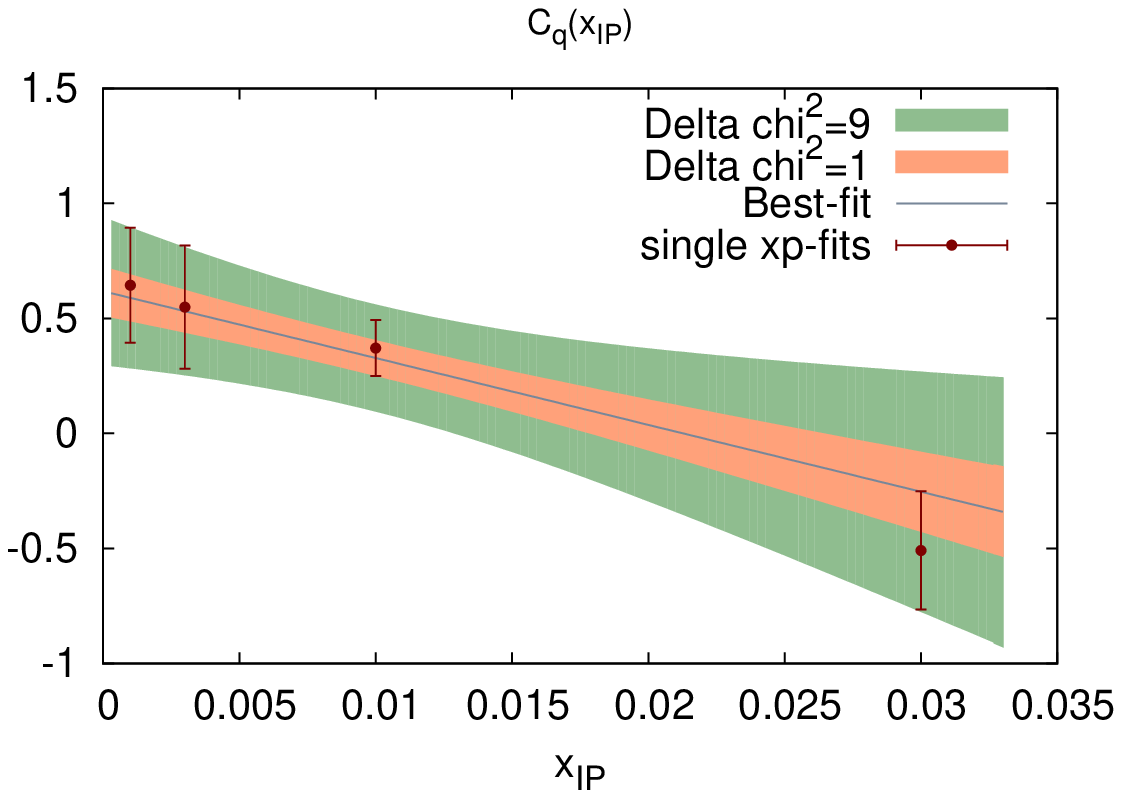}
\end{minipage} 
\vspace*{-0.5cm}
\end{center}
\caption{\small{Parameters as a function of $x_\pom$. Red dots are the results from pQCD fits at fixed $x_\pom$.
The grey line are best-fit prediction from $x_\pom$-combined fit. The bands represent the propagation 
of experimental uncertainties by using the Hessian method~\cite{CTEQ}.}}
\label{pars}
\end{figure}

\noindent
The alternative method we propose is instead inspired by the factorisation theorem 
\cite{DDISfactorisation} for diffractive DIS itself. 
The latter states that factorisation holds at fixed 
values of $x_\pom$ and $t$ so that the parton content described by
$f_i^{D}$ is uniquely fixed by the kinematics of the outgoing proton
and it is in principle different for different values of $x_\pom$ (and $t$, eventually). 
This idea is realised in practice by performing a series of pQCD fits 
at fixed values of $x_\pom$ with a common initial condition 
controlled by a set of parameters $\{p_i\}$. 
This procedure guides us to infere the approximate dependence of parameters $\{p_i\}$ on $x_\pom$
allowing the construction of initial condition in the $\{\beta,x_\pom\}$ space
to be used in a global fit, without any further model dependent assumption.
For the fits at fixed $x_\pom$ we choose the following singlet and gluon distributions 
at the arbitrary scale $Q_0^2$:
\begin{eqnarray}
\label{ic}
\beta \; \Sigma({\beta,Q_0^2}) &=& A_q \; \beta^{B_q} \; (1-\beta)^{C_q} \; e^{-\frac{0.01}{1-z}}\,,\nonumber\\
\beta \; g ({\beta,Q_0^2}) &=& A_g \; e^{-\frac{0.01}{1-z}}\,,\nonumber
\end{eqnarray}
which have four free parameters. 
We further assume that all lights quark distributions are equal to each other.
The exponential dumping exponential factor allows more freedom in the variation of the parameters $C_q$ at large $\beta$ 
and we choose the gluon distribution to be a simply a costant at $Q_0^2$~\cite{H1LRG06}. 
Such distributions are then evolved with the \texttt{QCDNUM17}~\cite{QCDNUM17} program  
within a fixed flavour number scheme to next-to-leading order accuracy.
Heavy flavours contributions are taken into account in the general massive 
scheme.
The convolution engine of \texttt{QCDNUM17} is used to obtain 
$F_2^{D(3)}$ and $F_L^{D(3)}$ structure functions at next-to-leading order 
which are then minimised against H1 data~\cite{H1LRG06}. In order to avoid the resonance region, 
a cut on the invariant mass
of the hadronic system $X$ is applied, $\mx^2\ge 4  \, \mbox{GeV}^2$.
Fixed $x_\pom$-fit results are sensitive 
to the choice of the mininum $Q^2$ value of data to be included in the fits.
The inclusion in the fits of data for which $Q^2 < 8.5 \, \mbox{GeV}^2$ in general worsens the 
$\chi^2$ and induce large fluctuation in the gluon distribution.
This effect has been already noticed in Ref.~\cite{H1LRG06} and avoided 
by including in the fit only data for which $Q^2 \ge 8.5 \, \mbox{GeV}^2$. 
The same strategy will be adopted here. Good quality fits have been obtained 
with the common initial condition for all values of $x_\pom$-bins~\cite{ourwork}.
The dependence of the parameters (as returned by the fits at fixed $x_\pom$)
on $x_\pom$ is shown in Fig.~[\ref{pars}]. 
Red dots are the results from pQCD fits at fixed $x_\pom$.  
The singlet normalisation $A_q$ behaves as an inverse power of $x_\pom$. In order to improve 
the description at higher $x_\pom$, however, an additional term is also included:   
\begin{equation}
\label{aq}
A_q(x_\pom) = A_{q,0} \; (x_\pom)^{A_{q,1}} \; (1-x_\pom)^{A_{q,2}} \,.\nonumber
\end{equation}
The gluon normalisation is compatible with a single inverse power behaviour of the type:
\begin{equation}
\label{ag}
A_g(x_\pom) = A_{g,0} \; (x_\pom)^{A_{g,1}}\,.\nonumber
\end{equation}
The coefficients $B_q$ and $C_q$ which control the $\beta$-shape of the singlet distribution 
are well described by:
\begin{eqnarray}
B_q(x_\pom) &=& B_{q,0} + B_{q,1} \, x_\pom \,, \nonumber\\
C_q(x_\pom) &=& C_{q,0} + C_{q,1} \, x_\pom \,. \nonumber
\label{bcq}
\end{eqnarray}
The following generalised initial condition 
\begin{eqnarray}
\label{ic2}
\beta \; \Sigma({\beta,Q_0^2,x_\pom}) &=& A_q(x_\pom) 
\; \beta^{B_q(x_\pom)} \; (1-\beta)^{C_q(x_\pom)} \; e^{-\frac{0.01}{1-z}}\,,\nonumber\\
\beta \; g ({\beta,Q_0^2,x_\pom}) &=& A_g(x_\pom) \; e^{-\frac{0.01}{1-z}}\,,\nonumber
\end{eqnarray}
is then used to perform a $x_\pom$-bin combined fit.
The combined fit has nine free parameters. Following the procedure
described in Ref.~\cite{PZ}, to each systematic errors quoted in the 
experimental analysis is assigned a free systematic parameters which 
is then minimised in the fit along with theory parameters. 
As for the single-$x_\pom$ fits, 
only data points for which  $\mx^2\ge 4 \, \mbox{GeV}^2$ 
and $Q^2 \ge 8.5 \, \mbox{GeV}^2$  are included in the fit.
The latter has an appreciable sensitivity on the scale $Q_0^2$
due to the relative stiffness of the initial condition.
The choice of $Q_0^2$ is then optimised performing a scan which 
gives the best $\chi^2$ value for $Q_0^2=2.3 \, \mbox{GeV}^2$. 
\begin{figure}[t]
\begin{center}
\begin{minipage}{16pc}
\includegraphics[width=16pc]{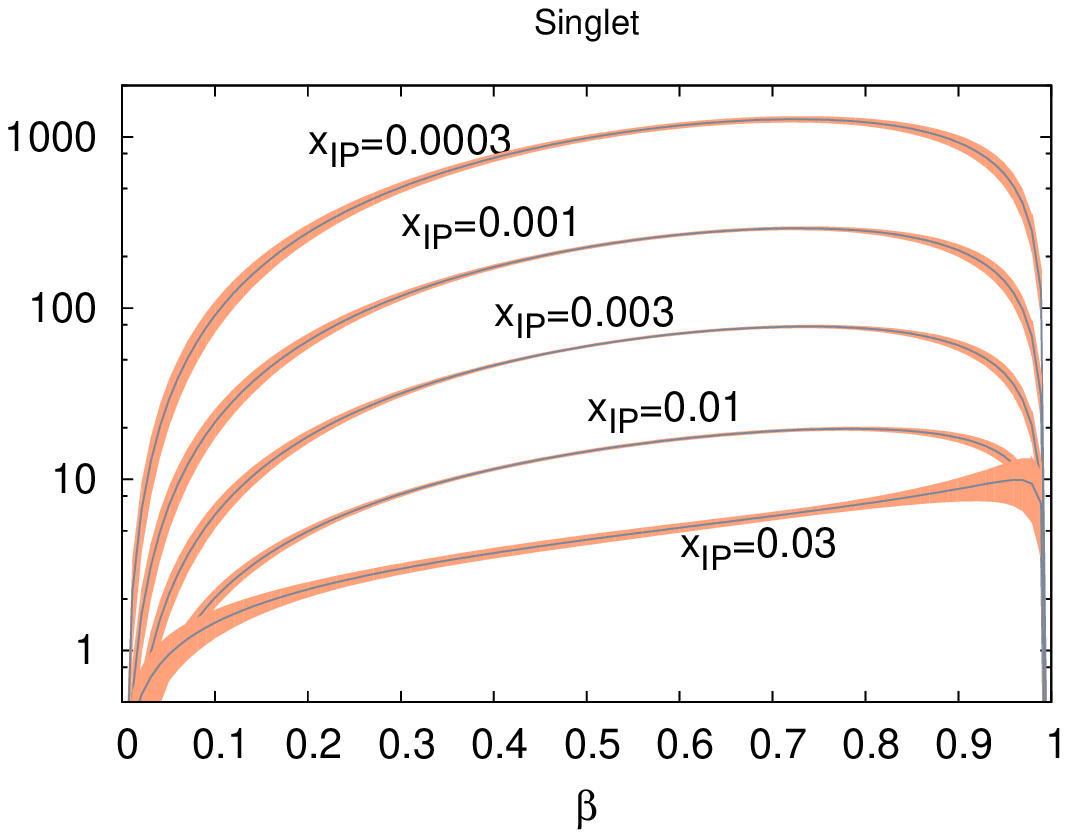}
\vspace*{-0.5cm}
\end{minipage}
\begin{minipage}{16pc}
\includegraphics[width=16pc]{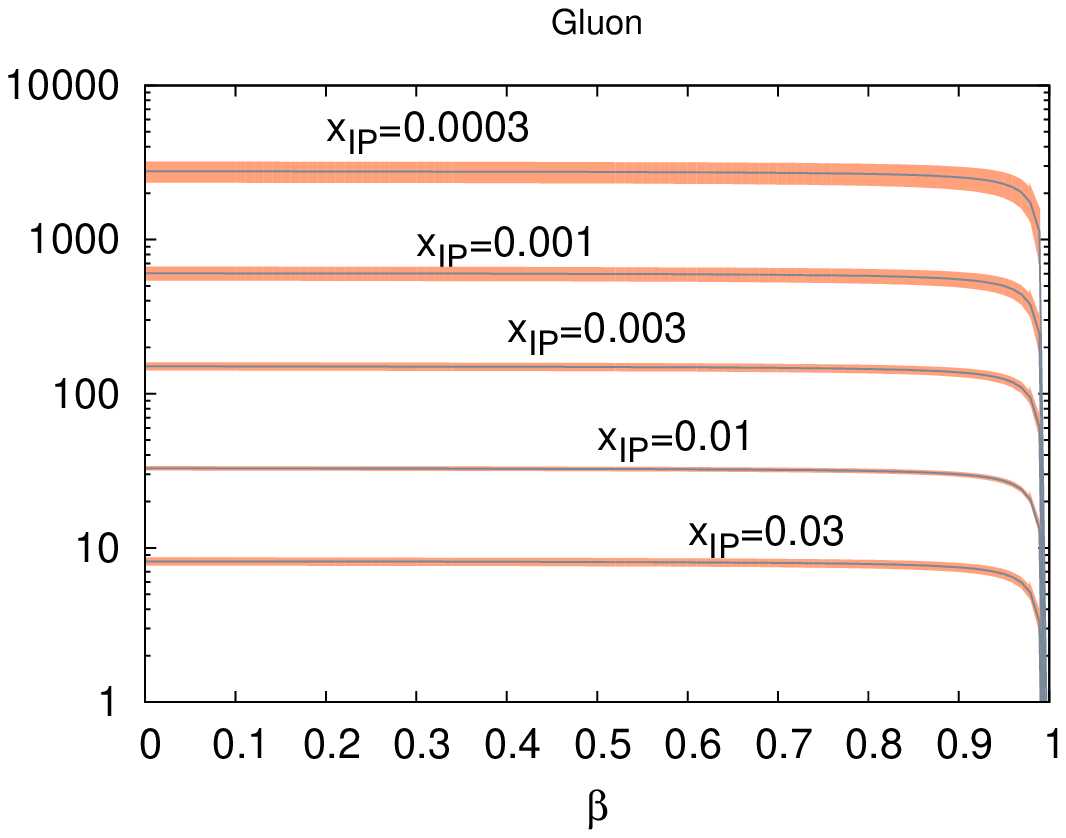}
\vspace*{-0.8cm}
\end{minipage}
\end{center}
\vspace{-0.5cm}
\caption{\small{Singlet and gluon initial condition at $Q_0^2$ as a function of $\beta$ for different $x_\pom$-values.
The bands represent the propagation of experimental uncertainties by using the Hessian method~\cite{CTEQ}.}}
\label{icfig}
\end{figure}
The best fit returns a $\chi^2=166$ for 182 degrees of freedom
which is of comparable quality as the one presented in Ref.~\cite{H1LRG06}. 
The initial condition allows the singlet and gluon 
normalisation, $A_q$ and $A_g$ respectively, to have a different power behaviour. 
It is therefore interesting to notice that if the condition $A_{q,1}=A_{g,1}$
is enforced, this results in a global increase of the $\chi^2$ to 171 units for 183 degree of freedom. 
If one further neglects the $x_\pom$-dependence of $B_q$ and $C_q$ by setting 
$B_{q,1}=C_{q,1}=0$ the $\chi^2$ increases to 188 units for 185 degree of freedom. 
This is an \textsl{a posteriori} confirmation that not only diffractive parton distributions 
change their magnitude versus $x_\pom$ but also that a modulation in their $\beta$-shape
(for the singlet, in this case) is necessary to better fit the data. 
The initial condition at $Q_0^2=2.3 \, \mbox{GeV}^2$ as a function of $\beta$ 
for different values of $x_\pom$ are shown in Fig.~[\ref{icfig}].

\section{Conclusions}
\noindent
We have outlined a new method to extract diffractive PDF's inspired by the factorisation theorem for diffractive DIS.  
From a series of pQCD fits at fixed $x_\pom$ we were able to infere the dependence of 
parameters on such a variable and this allowed us to construct 
a generalised initial condition without assuming neither proton vertex factorisation 
nor the existence of a series of Regge trajectories. 
The best-fit returns a $\chi^2$/d.o.f. close to unity, as the Regge-based pQCD
fit of Ref.~\cite{H1LRG06}, but in our opinion
the new procedure treats the non-perturbative $x_\pom$-dependence of the 
cross-section in a controlled and less model dependent way and
it might be capable (or even necessary) to fully exploit the expected improved precision of HERA-II data~\cite{H1FPS,H1LRG,H1VFPS}.\\

\noindent
F.A.C. would like to thank Dimitri Colferai and Ian Brock  
for their kind invitation to DIS12 Conference and the University of Bonn for support.   
L.F. is supported by the Fonds National de la Recherche Scientifique Belge (FNRS). 

{\raggedright
\begin{footnotesize}

\end{footnotesize}
}


\end{document}